# Chapter 3. Perception of the Environment

Author: Martin Drasar

**Abstract:** This chapter discusses the intricacies of cybersecurity agents' perception. It addresses the complexity of perception and illuminates how perception is shaping and influencing the decision-making process. It then explores the necessary considerations when crafting the world representation and discusses the power and bandwidth constraints of perception and the underlying issues of AICA's trust in perception. On these foundations, it provides the reader with a guide to developing perception models for AICA, discussing the trade-offs of each objective state approximation. The guide is written in the context of the CYST cybersecurity simulation engine, which aims to closely model cybersecurity interactions and can be used as a basis for developing AICA. Because CYST is freely available, the reader is welcome to try implementing and evaluating the proposed methods for themselves.

## 1 Background

Perception is a critical component of AICA and one of the few that cannot be omitted. Perception provides information about the environment, communicates the results of the agent's actions, and shapes and influences the agent's reasoning. While it may be possible to consider only the raw data gathered from sensors as the perception, this narrow view does not appreciate the complexity involved and only defers the issues of percept processing to other parts of AICA, such as the decision-making engine.

Perception in AICA is as multifaceted concept as it is in biological systems. Even though the artificial systems have the benefit of not being required to copy nature, many of the constraints and drivers are universal. The raw percepts or stimuli go through a lot of preprocessing and transformations before they can be subjected to the reason. Consider the optical illusion in Figure 1. Our brain is hardwired to identify real-world objects, so we get thrown off because they are not there. Moreover, it takes actual willpower to treat this image as just an image. The perception mechanisms shape how we think about our environment, and the same goes for AICA.

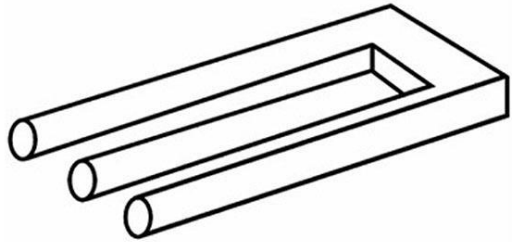

*Figure 1: An optical illusion.*

There are multiple ways to conceptualize the perception in AICA. One possible way is in the context of the DIKW pyramid, which conceptualizes the relation between data, information, knowledge, and wisdom (Ackoff, 1989). This is depicted in Figure 2, where the perception occupies the two lower tiers of the pyramid (data and information) but can sometimes venture up to the knowledge tier due to its close relation with AICA's world model. Another way we will adopt in this chapter is a pipeline, as shown in Figure 3, consisting of four main parts: physical sensors, logical sensors, transformers, and the world representation.

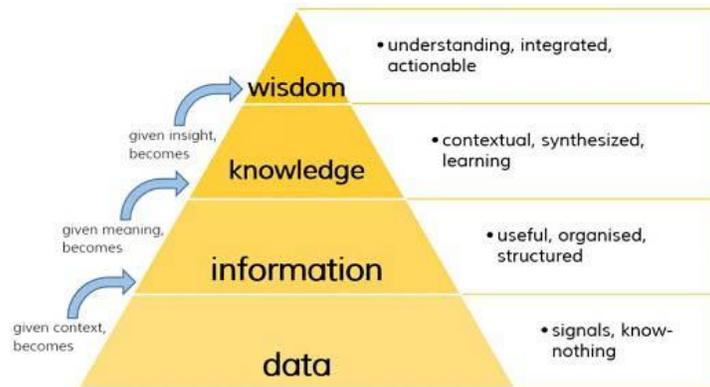

*Figure 2: DIKW pyramid (Baldasarre, 2017)*

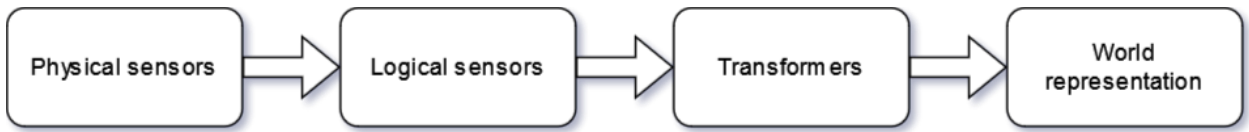

*Figure 3: A simple perception pipeline.*

**Physical sensors:** are primarily out of the scope of AICA. Physical sensors process non-virtual stimuli reaching the agent from the environment. Each of these sensors has specific operation capabilities, requirements, and physical domain, but they all share the need for power. Therefore, AICA using physical sensors must very carefully manage its power envelope.

**Physical sensor examples:** Temperature or pressure sensors and noise detectors could be employed by AICA tasked with maintaining physical security inside a building. Perimeter sensors could be used in outside deployment. Gyroscopes and lidars may be used within the context of unmanned vehicles.

**Logical sensors:** in the context of this chapter, they are understood as a counterpart to the physical ones. That is any source of data that rests within the software. A vast range of data can be fed to AICA in this way. Ranging from its internal state measurements, host diagnostics, and network measurements to open-source intelligence readings and even news feed. The only common attribute of this data is that there is nothing in common. The data provided by logical sensors is heterogeneous, with many dimensions, and can potentially require a large bandwidth to process. These attributes go counter to the current reinforcement learning algorithms, so there is a need for data reduction.

**Logical sensor examples:** Reading of running processes to gather information about the state of AICA and the infrastructure it operates in. Network probe to gather information about traffic within a guarded infrastructure. A periodic download of the CVE (MITRE) database to provide updates to AICA's knowledge base.

**Transformers:** provide means to reduce data complexity, dimensionality, and size. They ensure the move from the data tier of the DKIW pyramid up to the information tier. They can provide additional semantics to the data and serve as a heuristic that offloads a part of the logic that we do not want the ML algorithms to discover. There are many different types of processors, arguably more than types of data. The selection of transformers ultimately dictates how an agent perceives the environment and how it can reason about it.

**Transformer examples:** Statistical aggregation and transformation of observed network traffic (from packet traces to flows). Anomaly detection (from flows to events). Application of ML-driven tools (from events to patterns).

**World representation:** is AICA's representation of itself and of the environment it operates in. A model of the world as it is being perceived. It is the foundation on which the agent chooses its actions and against which their impact is evaluated. Currently, there exist no firm guidelines for the design of state representation. If anything, it is considered an art by some because the representation influences which algorithms can be used, how demanding the agent's training will be, and ultimately, what the agent can achieve.

Even though a pipeline is a fitting and easy-to-grasp concept, it gives an illusion of serial data processing. However, the sensors are usually independent, and the same mostly holds for transformers. As the data is being processed in parallel, delays, time skews, and interval differences are bound to happen, as illustrated in Figure 4. The impact of these irregularities strongly depends on the agent's mode of operation and choice of algorithms. Passive observing agents are largely unaffected because they can evaluate snapshots of the world state as the data comes in. However, for active agents, this de-serialization can impact AICA's efficiency by providing only partial observations over a longer time, thus impacting both learning and acting.

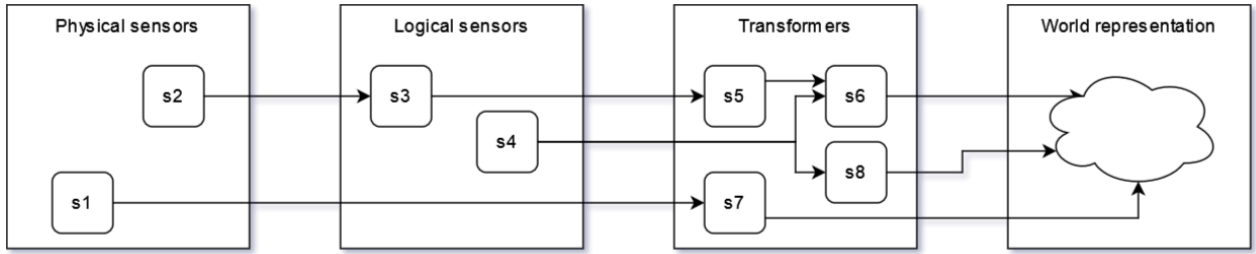

*Figure 4: A complex perception pipeline.*

This chapter addresses the complexity surrounding the perception and provides readers with guidelines and state-of-the-art examples. It does not present definite solutions, as many of these problems are still open and subject to research, but all the presented approaches have either been peer-reviewed or tested in attempts to develop a functional AICA.

## 2 From percepts to the world representation

AICAs will always operate in a partially observable environment. In fact, the observations provided by sensors will usually cover only a sliver of the environment. AICA will not observe, among other things, triggers that make other actors behave the way they do. Therefore, to enable rational and sensible actions, AICA must construct its belief state to be as close to the objective world state as possible. Through a sensible world representation, perception can go a long way to enable AICA to do just that. Conversely, choosing a suboptimal world representation will widen the gap between the belief and the objective state.

The world representation, as constructed from observations, can be split into three categories depending on their complexity and expressive power – atomic, factored, and structured (Russel & Norvig, 2020). With the atomic representation, states are indivisible and without an internal structure. This is the equivalent of perception playing no role in shaping the agent's understanding and providing only raw inputs to the decision-making engine. With the factored representation, incoming percepts are processed and represented as collections of attributes. These attributes may be primary, where parts of raw inputs are given their semantics, or secondary, where raw inputs are transformed into higher-level representations encoding some knowledge. With structured representation, attributes also encode their relation to other attributes. Going from atomic, over factored to structured representation leads to a sharp increase in expressiveness, where the world representation can concisely describe a complex environment and its interactions. However, this increase in expressiveness causes an inevitable increase in complexity, impacting reasoning and learning. Real-world AICA thus may be forced to combine representations of all three categories, carefully balancing the upsides and downsides.

Orthogonal to complexity, nevertheless with a considerable impact on creating a sensible world representation, is the matter of how perception deals with time. While analog sensors may measure continuously and provide an uninterrupted stream of stimuli, perception in AICA is inevitably discrete, with processing being done in independent time slices (Russel & Norvig, 2020). Meanwhile, sensors are unlikely to be synchronized, and their readings (or transformations) arrive at various intervals. It is then bound to happen that percepts related to one event will be split between two or more time slices. This, in turn, can impact the decision-making because the responses to AICA's actions may be incomplete. Three strategies can be used to counter this effect: slice extension, multi-slice perception, and contextual perception. Slice extension, as the name suggests, extends the time frame when percepts are collected. The problem persists, but the frequency of occurrence decreases, and the impact could be considered acceptable at some point. The downside is that an acceptable interval may be long enough to hamper AICA's speed of reaction to the point of jeopardizing its mission. With multi-sliced perception, the percepts are sampled in parallel with different interval lengths. Perception then produces multiple state updates, and AICA needs to have a strategy to cope with that, either on the perception level or at the decision-making level. With contextual perception, the percepts are still sampled; however, 1 to N neighboring samples are inspected, and the completeness of percepts is evaluated in relation to AICA's actions. This approach is the most complex one, as it requires the perception to have a clear model of which percepts occur and when. As with the complexity issue above, real-world AICA will likely have to combine all three approaches, carefully balancing the trade-offs.

The last consideration when designing the perception mechanism of AICA is the distinction between active and passive sensors, which can also be viewed as a distinction between the pull and the push model. Active sensors (pull) gather percepts as a result of their interaction with the environment. Passive sensors (push) receive stimuli from the environment and do not exert control over when and how it happens. As such, active sensors can be set in such a way to diminish the impact of the aforementioned sampling issue. However, this usually comes with considerably increased power or bandwidth requirements, as mentioned in the following text.

# 3 Power and bandwidth constraints

A naïve wisdom would suggest that the more sensors and the more sensory inputs, the better. After all, every new sensor can shorten the gap between the world representation and the world's objective state. New sensors can provide new auxiliary readings, additional details to already present sensors, and even wholly new percepts enabling the AICA to understand the world around itself. However, with each added sensor, there is a trade-off (Theron, et al., 2020). On the pure hardware level, each enabled sensor equates to energy expenditure. Whether this is a problem is a matter of AICA's deployment. Large stationary installations would probably be unaffected; however, AICAs on mobile platforms, personnel, or autonomous devices will have a strict power envelope, and the decision which sensors to use and when will rest on many factors, which are also shared with sensors on the software level. These usually do not have such stringent power limitations; rather, their issue is bandwidth. With pull-based sensors, too short sampling rate or too broad data collection can easily overwhelm the ability of AICA to process and reflect on the data.

When designing an AICA, one has to balance several sensor properties and prioritize sensors providing maximum utility for AICA's operation.
- Sensors (or their percepts) should be ordered by their importance for the decision-making process. This entails understanding how the percepts are transformed into the world representation and how the representation influences the decision-making. This can either be achieved through methods of explainable AI or by extensive testing evaluating the importance of each sensor.
- If the hierarchy is established, a base set of sensors should be selected, and AICA should activate the rest on demand.

- Especially for power-constrained environments, there should be a strategy to limit sensor function with the smallest possible impact on decision making, e.g., turning off sensors, prolonging sampling intervals, switching from pull to push mode, etc.
- AICA's decision-making should also be fortified against sensor impairment or partial sensor subversion.

# 4 Trusting the perception

One common theme in the literature is that the sensors provide objective input to agents' systems. Whether they are physical or logical sensors, it is taken for granted that the percepts they are producing are forming the world representation that is a clear reflection of the objective world state. This, however, need not hold in deployment settings. In fact, unintentional or deliberate fault of sensors can widen the gap between AICA's belief and objective state so much that the actions of an agent will go contrary to its goals.

Attacks against physical sensors have been studied in the literature (Nasralla, García-Magariño, & Lloret, 2020) (Man, Li, & Gerdes, 2020). Recently, the interest has been in the area of autonomous cars (Yan, Xu, & Liu, 2016) (Liu & Park, 2021); however, for any potential AICA deployment where physical sensors may play a role, the same concepts apply. While the faults cannot be eliminated, there are ways to build fault tolerance into the system, namely into data acquisition and data processing. For data acquisition, that can take the form of active probing of the environment against a known baseline (Shoukry, Martin, Yona, Diggavi, & Srivastava, 2015). For data processing, a simple sensor redundancy, fault-tolerant approaches researched in the area of sensor networks, and others can be used (Modarez, Kiumarsi, Lewis, Frank, & Davoudi, 2020). Regardless of the chosen approach, there will be incurred costs stemming from the need to increase the number of physical sensors, both as a procurement cost and increased power envelope.

For attacks against the logical sensors, mostly the same holds as for the physical ones. Active probing and sensor redundancy can be employed with the same expected results; however, some measures may be unattainable. Consider the example of the AICA measuring the state of the machine it is on. If the adversary managed to hide itself via hijacking certain syscalls, no amount of sensor redundancy would help because ultimately, every probe or every query would end up calling said syscalls, and the adversary would remain hidden. In such a case, only indirect information may hint at the presence of an adversary. One may argue that the time when an adversary hijacks syscalls is the time when the machine is effectively lost, but the same principle applies in different scenarios, where there is only one ultimate source of information for logical sensors, which is susceptible to subversion.

Considering the previous paragraphs, the perception cannot and should not be fully trusted, and the possibility of its subversion should be taken into account, especially when AICA is being built as a resilient solution operating in an adversarial environment. However, the price to maximize the trust in perception may be too high, and alternative solutions may have to be employed. Aside from fault-tolerant decision making, which is explored later in this book, multi-agent setups of AICA allow for perception sharing. In such a case, the setup can be considered a sensor network, and all the approaches, issues and limitations apply.

Finally, the perception may not only be a victim of an external adversary but also of wrong expectations. The purpose of perception is mapping sensory inputs to possible real-world states, with the key word being "possible" here. Most sensors come with expectations about the domain of possible values or their combination. However, the vast history of program faults caused by unexpected inputs should be treated as a cautionary tale. With physical sensors, the domain of percepts is bound by physical laws, but with logical sensors, all bets are off. That is why we have a lot of provably secure bridges and not many provably secure programs.

# 5   Developing a perception model for AICA

Perception models, i.e., world representations and associated transformations, are being extensively researched in the areas of autonomous cars, planes, and robots, where correct processing of environmental stimuli is of paramount importance. A similar situation is in natural language processing, where approaches to word embedding for encoding semantic similarity between words can also be considered a perception model for natural languages (Mikolov, Corrado, Chen, & Dean, 2013).

However, perception models for fully virtual entities like AICA are not extensively researched. As mentioned earlier, sensory inputs are treated as objective, and the creator of such a virtual entity is left to develop the perception model on their own, despite there being no solid guidance in the literature. Even the seminal work of Russel and Norwig, which provides probably the most complete exploration of the field of artificial intelligence, only skirts over this topic and rather focuses on the transformations of visual and physical stimuli. However, the book at least presents three essential properties which a good world representation should have (Russel & Norvig, 2020):

- it contains enough information for the agent to make good decisions,
- it is structured so that it can be updated efficiently,
- it is natural in the sense that it corresponds to the real world.

These are essential properties but not as easy to use as a starting point.

Modeling the cybersecurity domain, i.e., creating a perception model that is a good representation of the environment and satisfies the three properties above, is not an easy task. Unlike the scenarios that are being used across the literature, the cybersecurity domain in its entirety is highly dynamic, ever-expanding, and complex. The model has to reflect this to provide actionable information to the agent. Nevertheless, with such complexity, one can easily run into the so-called curse of dimensionality, when the total number of states that an agent can encounter is only a tiny fraction of states that exist in a world representation. And the agent would be wasting scarce resources to try and work with it. At the same time, it is not possible to simply resort to methods reducing the dimensionality of the representation, such as low-dimensional embedding via unsupervised learning (Saul & Roweis, 2003) or principal component analysis. While these methods are perfectly applicable in a technical sense, lowering the dimension count risk going counter to one of the aforementioned properties – that the representation is natural. If AICAs are ever to be used as a replacement for human cybersecurity experts or trusted with control over infrastructure, a key requirement will be full auditability in the form of explainable AI. However, if sensor inputs are non-linearly transformed into compact representation, AICAs and humans lose a shared vocabulary for explanation.

Currently, the only way to create satisfactory perception models is to handcraft them together with required heuristics (transformations) and painstakingly evaluate their efficiency. The author is aware of research in the area of unsupervised dimensionality reduction, which preserves explainability; however, that research is still in too early phase to be useful to the reader.

As there do not seem to be guidelines for creating perception models in an area as complex as cybersecurity, the following text will present a couple of use cases, which should help readers gain insights useful for building their own models. These use cases were taken from real-life attempts to create autonomous attackers driven by reinforcement learning algorithms.

Each of these use cases was realized within the CYST cybersecurity simulation engine, which is, to our knowledge, the currently most complex cybersecurity simulator that is freely available. (Drašar, Moskal, Yang, & Zaťko, 2020) CYST is a multi-agent discrete-event simulator based on message passing and tailored for cybersecurity applications. Given its complexity, only the parts relevant to the topic of perception are introduced in the following text. However, this chapter is accompanied by a code repository where the presented use cases are implemented, and readers are welcome to try and tinker with the ideas presented here.

## 5.1  Environment - the objective reality

The environment observed by the AICA is the environment simulated by CYST and defined by its simulation model. To minimize the cognitive load on the reader, this text uses only the bare minimum needed to execute and understand the presented use cases. However, if the reader is so inclined, they can further explore the simulation model in the relevant paper or the CYST's documentation (Drašar, CYST, 2022).

The infrastructure where AICA resides consists of simulated machines on which services are running. These machines are connected via a simulated network that replicates an ethernet network without networking details. The network is partitioned utilizing active network devices called routers. AICA is just one of the services running on one or more simulated machines. AICA communicates with or influences the environment through messages. These messages are also the only mechanism through which AICA can observe the environment.

The messages used in CYST come in two types: requests and responses. One request-response pair represents an entire exchange related to one AICA's action. The fragmentation related to, e.g., packets or even TCP sessions, is treated as an implementation detail; thus, the perception is fully realized through observing one response to each request. Messages are a collection of attributes, some of whom have a factually finite domain, some have a technically finite domain, and the domain for some is infinite. The following table summarizes the attributes and their function:

| Message | |
|---|---|
| id | unique identifier of a message. The id is the same for request and response in a pair. |
| type | request or response |
| src_ip, dst_ip | source and destination IP addresses of the message (IPv4 or IPv6) |

| src_service, dst_service | source and destination service (in simulation treated as a string, technically a port number) |
|---|---|
| ttl | message time to live (used to prevent routing cycles) |
| metadata | observable statistical properties, such as packet count, flow length, etc. |
| authentication/authorization | authentication or authorization token (multi-factor authentication intricacies are purposefully omitted) |
| session | the persistent connection between two services |

| Request (in addition to all Message attributes) | |
|---|---|
| action | an effect that AICA wants to achieve (for the purpose of this text, a string from a finite domain, otherwise a much more complicated structure) |

| Response (in addition to all Message attributes) | |
|---|---|
| status | structured description of the effect of the request. Contains origin (network, node, service, system), value (success, failure, error), and detail (an enumeration of possible values). |
| content | currently, unstructured data sent in response. |

| Session | |
|---|---|
| start | a tuple containing an originating IP address and a service of the session |
| end | a tuple containing a destination IP address and a service of the session |

These attributes are the variables that AICA can observe for the purpose of this text. The number of variables is higher within the CYST simulation, but these were omitted for clarity as the added complexity does not affect the proposed approaches. Also, despite the previously expressed concern about trust in perception, the presented use case treats all these observed attributes as trustworthy and reflecting the objective state, because CYST does not currently support fabrication of wrong percepts.

The following text presents several potential approximations of the objective state, which is perceived from the attributes of incoming responses. These approximations are largely independent, and their ordering rather reflects a thought process when developing the perception model than some kind of hierarchy.

## 5.2 First approximation - taking inputs verbatim
The first and probably the most straightforward way to represent the objective state is based on responses being the only percept that the AICA has. The world representation is constructed as a set of all possible response values.

**Size:** The size is $2^n$, where n is the number of bits in each response. If we take a compact representation of the response structure above (and give ourselves a bit of leeway in limiting the infinite domain attributes and set the strings at most 256 bits long), we will reach the n over 1500.
**Pros:** This representation is very easy to make. Just take the incoming response and pass it to the decision-making engine to process.

**Cons:** This is a clear case of the curse of dimensionality. States that are to be encountered during an AICA's run will represent only a minuscule portion of the entire state space, and the burden of data filtering and turning it into a reasonable belief state will be left to the decision-making engine, which will have to expend disproportional amount of energy.

## 5.3 Second approximation - elimination of (semi)static observations

As mentioned before, the number of states that could effectively be encountered is disproportionate to the size of the world representation. One of the reasons is that many observations are static or semi-static within the context of AICA's operation. Consider the type of message. It can be either a request or a response; however, the perception only processes the responses. This attribute is static and can be freely omitted without any loss of precision. The same goes for source IP addresses and services of both message and its session, as these are fixed for the AICA. Destination IP addresses can be considered semi-static if all AICAs activities happen within specific subnets. In such a case, it is not necessary to process the entire range of IP addresses, and only a subset can be a basis for world representation.

**Size:** The size is still $2^n$, where n is the number of non-static bits in each response.
**Pros:** This representation is still as easy as the first approximation to make and requires only limited analysis.
**Cons:** The actual reduction in world representation size depends on the nature of observations, and there is no easy way to specify a fixed upper bound.

## 5.4 Approximation detour - the interplay between request and responses

The selection of actions is not a responsibility of the perception, as it belongs to the decision-making engine. However, unlike many scenarios that can be seen in the literature, in cybersecurity, an action may have a similar complexity as a response. That is, not some tightly packed domain or one or more real numbers, but a complex structure dependent on the observed percepts. AICA thus must be able to use the data from the observation and must be able to use them accordingly.

Approximations of the objective state may reduce precision, especially the ones in the following text. Yet, AICA's actions may require precise attributes for their correct execution. Therefore, any lossy approximation or transformation must be accompanied by supplementary data to enable the reconstruction of the attributes within the decision-making engine.

## 5.5 Third approximation - indexing of large domains

Among the attributes in the responses, some expand the world representation unreasonably, at least considering the total number of states the AICA can encounter. For such large domains, it is better to keep a dictionary of encountered values and map them to an index that is used in the world representation.

**Size:** This approximation enables almost arbitrary size reduction of the world representation by specifying a fixed index size.
**Pros:** The size reduction does not come with a loss of information and benefits larger domains more. The approximation is still comparatively easy to implement. Using a fixed index with a reasonable eviction strategy can enable AICA to forget superfluous observations.

**Cons:** Using the index can hamper the transferability of the algorithms because the mappings of attribute values may not be static. This non-static property can also harm the learning algorithms, where a change in mapping between runs may lead to wrong transition function inference. Fixed index sizes risk unintended consequences in case of overflow. This is further exacerbated if AICA is trained in diverse and fluctuating environments, where diversity of percepts will fuel the index overflow.

## 5.6 Fourth approximation - state restructuring

There is a distinct difference between the objective state as was described earlier, and the contents of responses. AICA's version of this objective state - its belief state - is pieced from small probes of request-response pairs. However, there is no reason why the perception should not be modeled closer to the objective state as it is being understood.

This is one of the possible versions of world representation:

| Machine | | |
|---|---|---|
| IP | Services | Sessions |

| Machine | | |
|---|---|---|
| IP | Services | Sessions |

…

The perception is centered around the information about possible targets. For each target, an IP address, the running services, and active sessions are retained. All this information can be index-mapped, especially the services, as their domain is finite, and many services are likely to be shared among different machines. There would probably be a limit on the number of machines the AICA had in its operating memory.

**Size:** is likely to be similar to the third approximation. In this approximation, information is only restructured and not necessarily changed.
**Pros:** this heuristic approach removes the burden of understanding the world from AICA's decision-making, and it can focus more on the higher-level strategic decisions. It also provides a more natural representation that AICA's operator can understand.
**Cons:** depending on the restructuring, this approximation can help or hinder the decision-making process. It is thus very dependent on the capabilities of the person doing the restructuring.

## 5.7 Fifth approximation - explicit activity history

Operations in the cybersecurity domain naturally have complex dependencies on past events. The decision-making process thus has to keep track of what was done by AICA, how the counterparty reacted, how the infrastructure evolved, and so on. While these considerations can be technically modeled as a k-order Markov process, the k would be very large.

Current decision-making algorithms tackle these dependencies, e.g., through the use of LSTM neural networks, Gated Recurrent Units, and similar. However, training and imprinting these memories to be correctly used over disjoint response-request pairs can be resource-consuming or currently infeasible.

The alternative is for perception to act as an explicit memory that is (partially) taking the role of decision-making processes. In the presented use case, this could mean adding new attributes to the world representation by means of also observing the requests. The potential representation for a service can then look like this:

| Service | | | | |
|---|---|---|---|---|
| Name | Version | Vulnerable | Exploitation attempts | Time since the last exploitation |

In this case, *name* and *version* are taken from responses, *vulnerable* is evaluated by consulting the list of vulnerable services (CVE or such), *exploitation attempts*, and *time* are taken from requests.

**Size:** each new attribute expands the world representation; however, this expansion can be limited by carefully choosing an appropriate domain.
**Pros:** this approach, which is the first strong application of transformers into the perception pipeline, provides several guarantees that the dependency on LSTM and such do not. The memory over which the decision is being made is explicit, precise, and does not rely on gradual imprinting into a neural network. This explicitness also supports better explainability.
**Cons:** Some important information may be hidden from the decision-making process if the attributes are not chosen carefully.

## 5.8 Sixth approximation - additional transformations within the perception

This final approximation is an umbrella one for every other conceivable transformation that can be added to the perception pipeline. In principle, each new transformation moves the logic away from the decision-making engine through heuristics application. The goal is to let the decision-making engine concentrate on high-level decisions while automating the things that are possible to be automated. Today, this approach seems the most viable one to achieve notable results.

# 6 Summary and conclusions

Perception is a key component of AICA, strongly shaping and influencing decision-making. This chapter introduced perception as a pipeline that acquires, transforms, and stores the raw percepts into a form that benefits the decision-making engine the most. The extent of this benefit depends on several important decisions taken when developing a perception model of AICA:
- What are the intended complexity and expressive power of the world representation?
- How should the perception deal with time?
- Should it be actively polling the percepts or waiting for their arrival?
- What power or bandwidth constraints are there for percepts' processing, and what is the importance of specific sensors?
- How can perception be trusted in the adversarial environment?

This chapter discussed these questions and presented trade-offs associated with various decisions. It then delved deeper into developing an actual perception model for AICA. Because the cybersecurity domain where AICA operates is much more complex than the traditional environments used in the literature, it introduced CYST, a cybersecurity simulation engine whose simulation model was used as an objective reality on which the world representation building approaches were demonstrated. In total, six approaches to approximating the objective state were presented, and their properties were explored:
- Passing the raw percepts to the decision-making engine.

- Eliminating (semi)static observations.
- Using indexing to eliminate the impact of percepts with large domains.
- Restructuring the world state to a form useful for the decision-making engine.
- Keeping an explicit activity history.
- Including additional transformations within perceptions.

Because these approaches were developed in the context of CYST, which is freely available, users are welcome to try implementing them and experiment with their implementation.